\documentclass[11pt]{article}

\usepackage[]{acl}

\usepackage{times}
\usepackage{latexsym}

\usepackage[T1]{fontenc}

\usepackage[utf8]{inputenc}

\usepackage{microtype}

\usepackage{inconsolata}

\usepackage{graphicx}

\usepackage[utf8]{inputenc}

\usepackage{cleveref}
\crefname{section}{§}{§§}
\Crefname{section}{§}{§§}

\title{GaVaMoE: Gaussian-Variational Gated Mixture of Experts for Explainable Recommendation}

\author{
\textbf{Fei Tang}\textsuperscript{$\diamondsuit$} \quad \quad
\textbf{Yongliang Shen}\textsuperscript{$\diamondsuit$} \quad \quad
\textbf{Hang Zhang}\textsuperscript{$\diamondsuit$} \quad \quad
\textbf{Zeqi Tan}\textsuperscript{$\diamondsuit$} \\
\textbf{Wenqi Zhang}\textsuperscript{$\diamondsuit$} \quad \quad
\textbf{Zhibiao Huang}\textsuperscript{$\blacklozenge$} \quad \quad
\textbf{Kaitao Song}\textsuperscript{$\heartsuit$} \quad \quad
\textbf{Weiming Lu}\textsuperscript{$\diamondsuit$} \quad \quad
\textbf{Yueting Zhuang}\textsuperscript{$\diamondsuit$} \\
\textsuperscript{$\diamondsuit$}Zhejiang University \\
\textsuperscript{$\heartsuit$}Microsoft Research Asia \\
\textsuperscript{$\blacklozenge$}Baidu Inc., China \\
\texttt{flysugar@zju.edu.cn} \quad \texttt{syl@zju.edu.cn} \quad \texttt{22451046@zju.edu.cn} \quad \texttt{zqtan@zju.edu.cn} \\
\texttt{zhangwenqi@zju.edu.cn} \quad \texttt{huangzhibiao@baidu.com} \quad \texttt{kaitaosong@microsoft.com} \\
\texttt{luwm@zju.edu.cn} \quad \texttt{yzhuang@zju.edu.cn}
}

\begin{document}
\maketitle
\begin{abstract}
Recent advances in large language models (LLMs) have enabled natural language explanations in recommendation systems.
However, current LLM-based explainable recommendation approaches face three critical limitations: inadequate modeling of user-item collaborative preferences, insufficient personalization in generated explanations, and poor performance with sparse user interactions. 
We present GaVaMoE, a Gaussian-Variational Gated Mixture of Experts framework that addresses these challenges through a novel two-stage architecture. GaVaMoE first employs a Variational Autoencoder (VAE) with Gaussian Mixture Model (GMM) to learn rich collaborative preference representations and cluster users with similar behaviors. These learned representations then inform a multi-gating mechanism that routes user-item pairs to specialized expert models for generating targeted explanations. This hierarchical design enables both effective preference modeling and personalized explanation generation, particularly benefiting scenarios with sparse interactions by leveraging user similarities. Experiments on three real-world datasets demonstrate that GaVaMoE significantly outperforms existing methods in explanation quality, personalization, and consistency, while maintaining robust performance for users with limited historical data\footnote{\url{https://github.com/sugarandgugu/GaVaMoE}.}.
\end{abstract}
\section{Introduction}
Recommendation systems are fundamental to modern digital platforms, helping users navigate vast information spaces through personalized suggestions. A critical challenge in these systems is providing clear explanations for recommendations, as explanations significantly impact user trust and decision-making \cite{tintarev2015explaining, zhang2019tutorial}. This raises a fundamental question: \textbf{How can we generate personalized, high-quality explanations that effectively communicate the reasoning behind recommendations?}

Recent advances in large language models (LLMs) have revolutionized natural language generation, presenting new opportunities for explainable recommendations. Several pioneering works have demonstrated LLMs' potential in generating contextual explanations. PEPLER \cite{li2023personalized} introduced GPT-2-based transfer learning with user-item IDs as prompts, while LLM2ER \cite{yang2024fine} enhanced this approach with personalized prompting. XRec \cite{ma2024xrec} further integrated collaborative filtering signals and user-item profiles, bridging traditional recommendation techniques with LLM capabilities.

\begin{figure}[]
    \centering
    \includegraphics[width=\columnwidth]{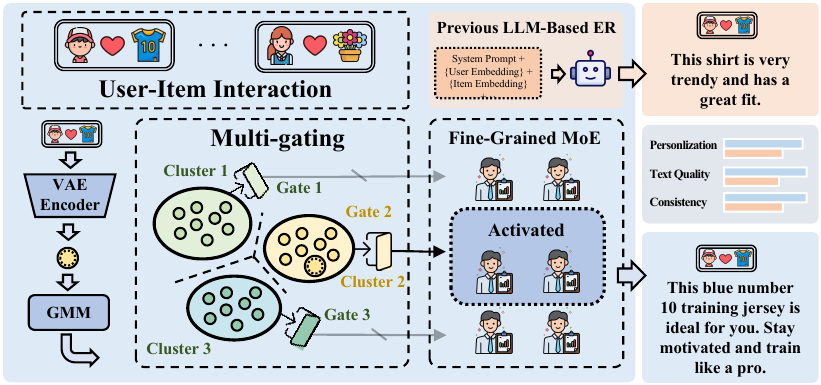} 
    \caption{Comparison between traditional LLM-based recommendation systems and GaVaMoE. (a) Traditional approaches directly map user-item IDs to LLM prompts for explanation generation. (b) GaVaMoE first encodes user-item interactions through VAE-GMM clustering, then routes them to specialized experts via multi-gating for personalized explanations.}
    \captionsetup{skip=0pt}
    \label{fig:motivation}
\end{figure}

However, current LLM-based explainable recommendation systems face three critical challenges: (1) \textit{Limited Collaborative Preference Modeling:} Existing approaches rely on simple concatenation of user-item IDs to model collaborative relationships. This direct mapping of discrete IDs to embeddings fails to capture the complex, non-linear relationships between users and items, resulting in suboptimal preference modeling. (2) \textit{Insufficient Personalization:} Generated explanations often lack user-specific nuance, producing generic recommendations that fail to reflect individual preferences and behaviors. This limitation stems from inadequate mechanisms for capturing and utilizing personalized user characteristics. (3) \textit{Poor Handling of Data Sparsity:} Current systems struggle with sparse user-item interactions, a common scenario in real-world applications. Without effective mechanisms to transfer knowledge between similar users or items, these approaches fail to generate meaningful explanations for users with limited historical data or for niche items.

To address these challenges, we propose GaVaMoE, a Gaussian-Variational Gated Mixture of Experts framework for generating personalized explanations in recommendation systems. Our key insight is that effective explanation generation requires both deep understanding of user preferences and specialized generation strategies for different user groups, as shown in Figure \ref{fig:motivation}. Building on this insight, GaVaMoE employs a two-stage approach: First, it leverages a Variational Autoencoder (VAE) combined with Gaussian Mixture Model (GMM) to learn rich collaborative preference representations and naturally cluster users with similar behaviors. These learned representations capture complex user-item relationships while providing a robust foundation for handling sparse data. Then, utilizing these learned representations and user clusters, a dynamic multi-gating mechanism intelligently routes user-item pairs to specialized expert models, each trained to generate explanations tailored to specific user preference patterns. This hierarchical design ensures both deep collaborative preference modeling and highly personalized explanation generation, effectively addressing the limitations of existing approaches.

Our contributions are threefold:

\begin{itemize}
\item We introduce a novel VAE-GMM architecture that effectively captures complex user-item collaborative preferences while naturally handling data sparsity through learned latent representations.
\item We develop a dynamic multi-gating mechanism that intelligently routes user-item pairs to specialized experts, enabling highly personalized explanations even for users with limited interaction history.
\item Through extensive experiments on three real-world datasets, we demonstrate that GaVaMoE significantly outperforms existing methods across multiple metrics, including explanation quality, personalization, and consistency.
\end{itemize}

\begin{figure*}[htbp]
    \centering
    \includegraphics[width=1\textwidth]{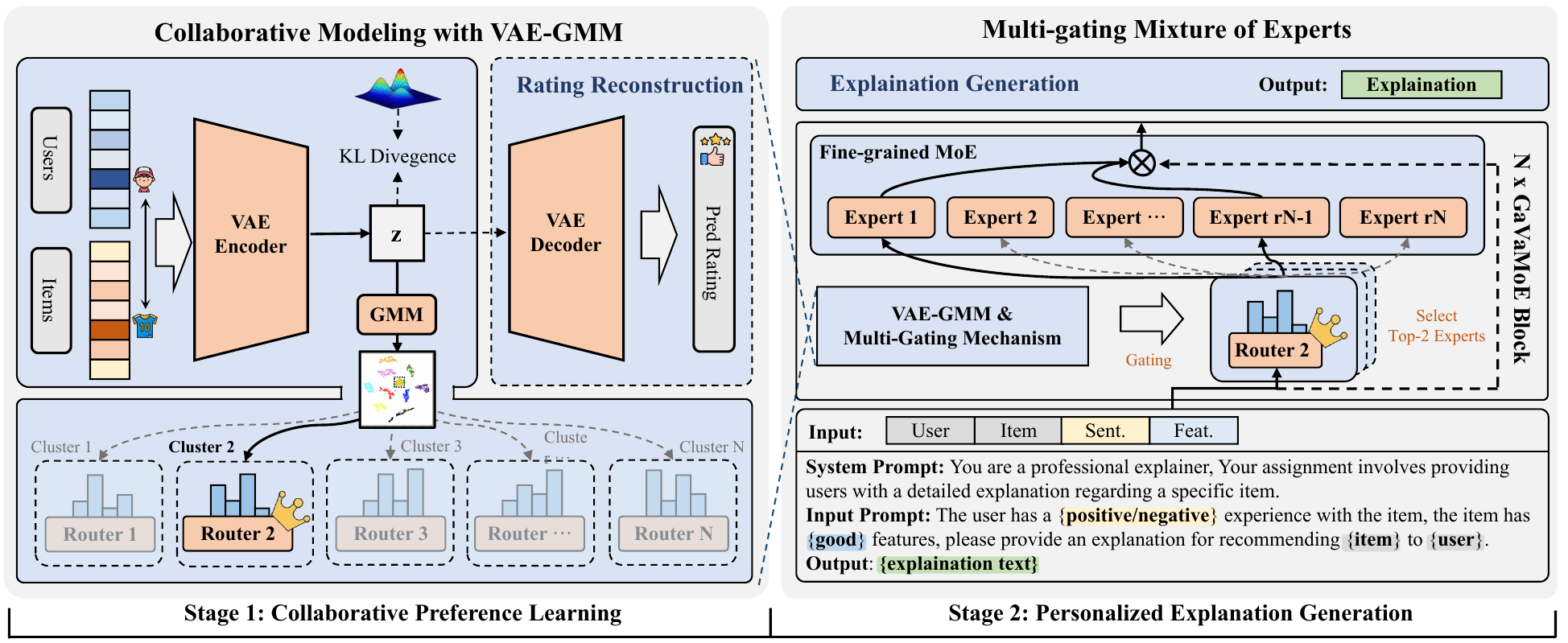} 
    \caption{Architecture overview of GaVaMoE. The framework consists of N stacked blocks, each containing: (1) a VAE-GMM module that learns collaborative preferences and clusters users with similar behaviors, and (2) a multi-gating mixture of experts that routes user-item pairs to specialized experts based on learned clusters. The model employs a two-stage training process, first optimizing collaborative preference learning and clustering, then fine-tuning for personalized explanation generation.}
    \label{fig:GaVaMoE}
\end{figure*}

\section{Method}

We present GaVaMoE, a framework that addresses the challenges of personalization and data sparsity in explainable recommendation through a two-stage approach. As illustrated in Figure \ref{fig:GaVaMoE}, GaVaMoE consists of two key components: (1) a VAE-GMM module for capturing deep collaborative preferences and clustering users (\cref{sec:3.2}), and (2) a multi-gating mixture of experts architecture for generating personalized explanations (\cref{sec:3.3}). We first formalize the task (\cref{sec:3.1}), then detail each component and their integration (\cref{sec:3.2}-\cref{sec:3.3}), and finally describe our training strategy (\cref{sec:3.4}).

\subsection{Task Formulation}
\label{sec:3.1}
The explainable recommendation task aims to generate both item recommendations and natural language explanations that reflect user preferences. Given a user-item pair $(u,i)$ where $u \in U$ is a user ID and $i \in I$ is an item ID, we observe a rating $r_{u,i} \in R_{>0}$ indicating user $u$'s preference for item $i$. Each item is associated with a set of features $f_{u,i} \in F$ (e.g., genre tags for movies) that characterize its attributes. Following \cite{yang2024fine}, we incorporate these features into the LLM prompt to enhance explanation generation. The input prompt undergoes tokenization and passes through the LLM's attention layers before reaching the routing component, ensuring effective encoding of semantic information.

Formally, given inputs $(u, i, r_{u,i}, f_{u,i})$, our model aims to generate a personalized explanation $\hat{e}_{u,i}$ that reflects both the user's preferences and the item's characteristics. The key challenge lies in capturing complex user-item collaborative relationships while generating explanations that are both personalized and informative, especially in scenarios with sparse user interactions.

\subsection{Collaborative Modeling with VAE-GMM}
\label{sec:3.2}
The foundation of GaVaMoE is a VAE-GMM module that jointly learns collaborative preferences and clusters users with similar behaviors. This component serves two crucial purposes: modeling complex user-item relationships and providing a principled basis for the multi-gating routing strategy.

\subsubsection{Learning Latent Preferences with VAE}

To effectively capture complex user-item collaborative relationships, we employ a VAE to learn compact and informative latent representations. Rather than directly modeling user-item interactions in their original sparse space, we map them to a dense latent space that encodes underlying preference patterns.

Formally, for each user-item pair $(u,i)$, the encoder $E$ maps their embeddings to a probabilistic latent space, parameterized by mean $\mu$ and log-variance $log(\sigma)^2$:
\begin{equation}
[\mu, log(\sigma)^2] = E(u, i)
\end{equation}

To enable effective gradient-based training while maintaining the probabilistic nature of our representations, we employ the reparameterization trick:
\begin{equation}
z = \mu + \varepsilon\sigma, \quad \varepsilon \sim N(0, I)
\end{equation}
This transformation allows us to sample from the learned distribution while maintaining differentiability. The sampled latent vector $z$ captures the user's preference patterns in a compact form.

The decoder $D$ then maps these latent preferences back to the rating space:
\begin{equation}
\hat{r}_{u,i} = D(z)
\end{equation}

This probabilistic encoding serves three key purposes: (1) it learns robust preference representations that capture underlying patterns in user behavior, (2) it provides a structured latent space that facilitates subsequent user clustering, and (3) it naturally handles sparse interactions by learning to generalize across similar preference patterns. These learned representations form the foundation for our preference-guided clustering and personalized explanation generation.

\subsubsection{User Clustering with GMM}
To enable personalized routing in our multi-gating mechanism, we extend the VAE with a Gaussian Mixture Model that clusters users based on their collaborative preferences. The GMM assumes that user-item embeddings follow a mixture of $K$ Gaussian distributions, defined as $P = \text{GMM}(\pi, \bar{\mu}, (\bar{\sigma})^2)$, where $\pi$ represents mixture weights, and $\bar{\mu}, (\bar{\sigma})^2$ are component-wise means and variances.

The joint distribution over observations $x$, latent variables $z$, and cluster assignments $c$ is:
\begin{equation}
p(x, z, c) = p(x|z)p(z|c)p(c)
\end{equation}
where:
\begin{align}
p(c) &= \text{Cat}(c|\pi) \\
p(z|c) &= \mathcal{N}(z|\bar{\mu}_c, (\bar{\sigma}_c)^2I) \\
p(x|z) &= \text{Ber}(x|\mu_x)
\end{align}

This formulation enables simultaneous learning of latent representations and cluster assignments through ELBO optimization (detailed in Appendix \ref{app:elbo}). The resulting clusters naturally correspond to gates in our multi-gating mechanism, providing a data-driven routing strategy that directs user-item pairs to appropriate expert models based on learned preference patterns.

\subsection{Multi-gating Mixture of Experts}
\label{sec:3.3}
Building on the learned user clusters from VAE-GMM, we design a multi-gating mixture of experts architecture that enables fine-grained personalization in explanation generation. Our approach extends traditional MoE by introducing cluster-aware routing and specialized expert decomposition.

\subsubsection{Multi-gating for Expert Selection}
We design a cluster-aware multi-gating mechanism to address two key challenges in explainable recommendation: (1) different user groups require distinct explanation strategies, and (2) users with similar preferences should share explanation patterns. 
The multi-gating mechanism leverages learned cluster information to route user-item pairs to appropriate experts. 

\paragraph{Cluster-based Gate Assignment} For an input with latent embedding $\mathbf{z}$, we first compute its soft assignment to each cluster using the posterior probability:

\begin{equation}
    \begin{aligned}
    \gamma_{c} = \frac{\pi_c \cdot \mathcal{N}(\mathbf{z} | \bar{\mu}_c,(\bar{\sigma}_c)^2)}
    {\sum_{c'=1}^{K} \pi_{c'} \cdot \mathcal{N}(\mathbf{z} | \bar{\mu}_{c'},(\bar{\sigma}_{c'})^2)} 
    \end{aligned}
    \label{eq:y}
\end{equation}
where $\pi_c$ is the prior probability of cluster $c$, and $\mathcal{N}(\mathbf{z} | \bar{\mu}_c,(\bar{\sigma}_c)^2)$ is the likelihood of $\mathbf{z}$ under cluster $c$'s Gaussian distribution.

\paragraph{Gate Selection and Routing} We establish a one-to-one correspondence between clusters and gates, where each cluster $c$ corresponds to a specific gate $G_c$ in our set of gates $G = {G_1, ..., G_K}$. The input is routed to the gate corresponding to its most probable cluster:
\begin{equation}
\bar{c} = argmax_c \gamma_{c}
\end{equation}

This cluster-aware routing design enables coherent explanation generation by directing users with similar preferences to the same gate, while leveraging collective knowledge within clusters to handle sparse user interactions. Each gate naturally develops specialized expertise for specific preference patterns, leading to more targeted and relevant explanations.

\subsubsection{Fine-grained Mixture of Experts}
Traditional MoE architectures face a fundamental trade-off between expert specialization and computational efficiency. While increasing the number of experts can enhance specialization, it leads to prohibitive computational costs \cite{dai2024deepseekmoe}. To address this challenge, we propose a fine-grained expert decomposition strategy that enables more precise modeling of explanation patterns while maintaining efficiency.

Instead of replicating entire feed-forward networks, we decompose each expert into smaller specialized units. Given an original expert with hidden dimension $d$, we reduce its dimension to $d/r$ while increasing the number of experts from $N$ to $rN$, where $r$ is the decomposition factor. This design maintains the total parameter count while enabling finer-grained specialization:
\begin{equation}
    N \cdot d = rN \cdot (d/r)
\end{equation}
For an input $x$ routed to gate $G_{\bar{c}}$, expert selection proceeds through a two-step process: First, we compute expert selection scores within the gate:
\begin{equation}
    G_{\bar{c}}(x)_i = \frac{e^{\mathbf{W} \cdot x}_i}{\sum_{j=1}^{rN} e^{\mathbf{W} \cdot x}_j}
\end{equation}
where $\mathbf{W}$ represents the gate-specific routing weights. Then, we select the top-k experts ($\Omega$) and combine their outputs:
\begin{equation}
    y = \sum_{i \in \Omega}^{rN} G_{\bar{c}}(x)_i \cdot E_i(x)
\end{equation}
This fine-grained design enables more precise modeling of explanation patterns by allowing multiple specialized experts to collaboratively generate explanations. The top-k routing strategy ensures computational efficiency while maintaining the benefits of expert specialization. When combined with our cluster-aware routing, this architecture enables GaVaMoE to generate highly personalized explanations by matching user preferences with appropriate combinations of specialized experts.
\subsection{Two-stage Training Objective}
\label{sec:3.4}

Training GaVaMoE requires careful coordination between collaborative preference learning and explanation generation. We adopt a two-stage training strategy that first establishes robust user representations and then leverages these for personalized explanation generation.

\subsubsection{Stage 1: Collaborative Preference Learning}

The first stage focuses on learning deep collaborative preferences through the VAE-GMM component. We optimize the Evidence Lower Bound (ELBO):
\begin{equation}
\begin{split}
\mathcal{L}_{\text{ELBO}}(x, \beta) &= \mathbb{E}_{q(z, c | x)} \left[ \log p(x | z) \right] \\
&\quad - \beta \cdot \text{KL}\left(q(z, c | x) || p(z, c)\right)
\end{split}
\end{equation}

This objective balances reconstruction accuracy with representation learning through two terms:
(1) The reconstruction term $\mathbb{E}_{q(z, c | x)} \left[ \log p(x | z) \right]$ ensures accurate modeling of user-item interactions
(2) The KL divergence term, weighted by $\beta$, encourages structured latent representations that facilitate clustering

\subsubsection{Stage 2: Personalized Explanation Generation}

Building on the learned representations, the second stage trains the multi-gating mixture of experts for explanation generation. We optimize a combined objective:
\begin{equation}
\mathcal{L}_{\text{total}} = \alpha \cdot \mathcal{L}_{\text{ELBO}} + (1 - \alpha) \cdot \mathcal{L}_{\text{explanation}}
\end{equation}
where $\mathcal{L}_{\text{explanation}}$ represents the negative log-likelihood of generating correct explanations, and $\alpha$ balances preference modeling with explanation quality. This formulation ensures that the model maintains accurate collaborative preference understanding while learning to generate personalized explanations.

This two-stage approach offers several benefits: it ensures stable learning of user preferences, enables effective knowledge transfer from preference modeling to explanation generation, and maintains the quality of both collaborative filtering and explanation generation components.

\section{Experiment}
\subsection{Experiment Setup}
To evaluate GaVaMoE, we used three diverse datasets: TripAdvisor\footnote{\url{https://www.tripadvisor.com}}, Amazon (Movies \& TV)\footnote{\url{http://jmcauley.ucsd.edu/data/amazon}}, and Yelp\footnote{\url{https://www.yelp.com/dataset}}. TripAdvisor focuses on travel reviews, Amazon on movie and TV ratings, and Yelp on local businesses. As shown in Table \ref{tab:Statistics of the Datasets}, Yelp is the largest and potentially most sparse, followed by Amazon and TripAdvisor. This dataset variety helps assess GaVaMoE's adaptability to different user behaviors and data sparsity. Each dataset was split into training, validation, and test sets (8:1:1). For more details, see Appendix \ref{sec:details_train}.
\begin{table}[]
\begin{tabular}{@{}lccc@{}}
\toprule
           & TripAdvisor & Amazon  & Yelp      \\ \midrule
\#users    & 9,765       & 7,506   & 27,147    \\
\#items    & 6,280       & 7,360   & 20,266    \\
\#records  & 320,023     & 441,783 & 1,293,247 \\
\#features & 5,069       & 5,399   & 7,340     \\ \bottomrule
\end{tabular}
\caption{Statistics of the Datasets}
\label{tab:Statistics of the Datasets}
\end{table}

\begin{table*}[]
\resizebox{\textwidth}{!}{
\begin{tabular}{cccccccccccccccccccccccccc}
\toprule
\multicolumn{1}{l}{} & \multicolumn{2}{c}{BLEU (\%)}    & \multicolumn{2}{c}{ROUGE (\%)}    & \multicolumn{2}{c}{Distinct (\%)} & \multicolumn{3}{c}{BERTScore}               \\ \midrule
Method               & BLEU-1          & BLEU-4         & ROUGE-1         & ROUGE-L         & Distinct-1      & Distinct-2      & BERTScore-P    & BERTScore-R    & BERTScore-F    \\ \midrule
\multicolumn{10}{c}{Amazon(Movie \& TV)}                                                                                                                                           \\ \midrule
PEPLER               & 12.564          & 0.991          & 14.005          & 10.816          & 17.134          & 61.190          & 0.369          & 0.362          & 0.364          \\
PETER                & 14.796          & 1.201          & 15.335          & 11.614          & 17.431          & 62.790          & 0.384          & 0.376          & 0.380          \\
PEVAE                & 16.349          & 1.503          & 19.255          & 15.853          & 18.440          & 60.610          & 0.349          & 0.340          & 0.345          \\
NETE                 & 14.965          & 1.152          & 16.470          & 12.982          & 14.503          & 47.312          & 0.344          & 0.344          & 0.344          \\
XRec                 & 20.259          & 1.730 & 24.971          & 17.737          & 20.370          & 64.444          & 0.409          & 0.398          & 0.401          \\
GaVaMoE              & \textbf{21.370} & \textbf{1.751} & \textbf{25.445} & \textbf{17.747} & \textbf{21.254} & \textbf{65.784} & \textbf{0.416} & \textbf{0.410} & \textbf{0.412} \\ \midrule
\multicolumn{10}{c}{TripAdvisor}                                                                                                                                                   \\ \midrule
PEPLER               & 13.392          & 0.965          & 15.382          & 12.078          & 18.124          & 64.923          & 0.345          & 0.303          & 0.343          \\
PETER                & 14.957          & 0.881          & 16.514          & 13.002          & 19.400          & 65.652 & 0.354          & 0.325          & 0.350          \\
PEVAE                & 15.705          & 1.401          & 18.675          & 15.411          & 18.182          & 66.414          & 0.369          & 0.363          & 0.368          \\
NETE                 & 13.847          & 1.186          & 14.891          & 11.880          & 14.854          & 48.431          & 0.300          & 0.293          & 0.295          \\
XRec                 & 19.642          & 1.650          & 23.672          & 16.871          & 21.113          & 65.332          & 0.386          & 0.380          & 0.384          \\
GaVaMoE              & \textbf{21.142} & \textbf{1.891} & \textbf{25.465} & \textbf{18.118} & \textbf{22.784} & \textbf{68.398} & \textbf{0.397} & \textbf{0.392} & \textbf{0.394} \\ \midrule
\multicolumn{10}{c}{Yelp}                                                                                                                                                          \\ \midrule
PEPLER               & 9.448           & 0.623          & 12.604          & 9.769           & 13.321          & 43.577          & 0.317          & 0.294          & 0.305          \\
PETER                & 8.072           & 0.574          & 12.965          & 9.850           & 14.240          & 52.891          & 0.298          & 0.271          & 0.281          \\
PEVAE                & 14.250          & 1.217          & 16.229          & 13.891          & 16.890          & 58.510          & 0.328          & 0.323          & 0.326          \\
NETE                 & 11.314          & 0.823          & 14.041          & 9.226           & 13.245          & 44.426          & 0.222          & 0.137          & 0.144          \\
XRec                 & 19.379          & 1.601          & 22.870          & 16.531          & 17.560          & 61.190          & 0.395          & 0.351          & 0.373          \\
GaVaMoE              & \textbf{19.586} & \textbf{1.616} & \textbf{23.259} & \textbf{17.416} & \textbf{19.483} & \textbf{63.847} & \textbf{0.406} & \textbf{0.374} & \textbf{0.381} \\ \bottomrule
\\
\end{tabular}
}
\caption{GaVaMoE uses a clustering parameter K (the number of clusters), which is set to 3 for TripAdvisor and Yelp, and 2 for Amazon, corresponding to their respective optimal number of gates. More details on the parameters of GaVaMoE can be referred to the Appendix B.1.}
\label{AllPerformance}
\end{table*}
\subsection{Results and Analysis}
\subsubsection{Main Results}
Table \ref{AllPerformance} compares explanation generation methods on TripAdvisor, Yelp, and Amazon datasets. GaVaMoE outperforms all baselines, including XRec, across all metrics. Specifically, it improves over XRec by 6.39\% on TripAdvisor, 4.46\% on Yelp, and 2.57\% on Amazon. GaVaMoE also achieves the highest BLEU-4 and ROUGE-L scores, with a BLEU-4 of 1.75\% on Amazon, surpassing XRec’s 1.73\%. Its performance reflects fluent explanation generation. Additionally, GaVaMoE shows improved consistency, with a BERTScore-F of 0.412 on Amazon, compared to XRec’s 0.401. On the sparse Yelp dataset, GaVaMoE’s BERTScore-F of 0.381 shows a 2.14\% improvement over XRec, indicating its robustness in sparse data. Finally, GaVaMoE generates more diverse explanations, with higher Distinct-1 and Distinct-2 scores on TripAdvisor (22.784\% and 68.398\%) compared to XRec (21.113\% and 65.332\%).

These improvements can be attributed to GaVaMoE's key components. The VAE captures nuanced user-item interactions for more accurate, personalized explanations. Enhanced by GMM, it refines user grouping based on preferences, which informs a multi-gating mechanism. This mechanism routes user-item pairs to the most suitable experts, ensuring targeted and relevant explanations. The combination of deep preference modeling, clustering, and expert routing allows GaVaMoE to generate higher-quality, personalized, and consistent explanations.

\begin{figure}[]
    \centering
    \includegraphics[width=\columnwidth]{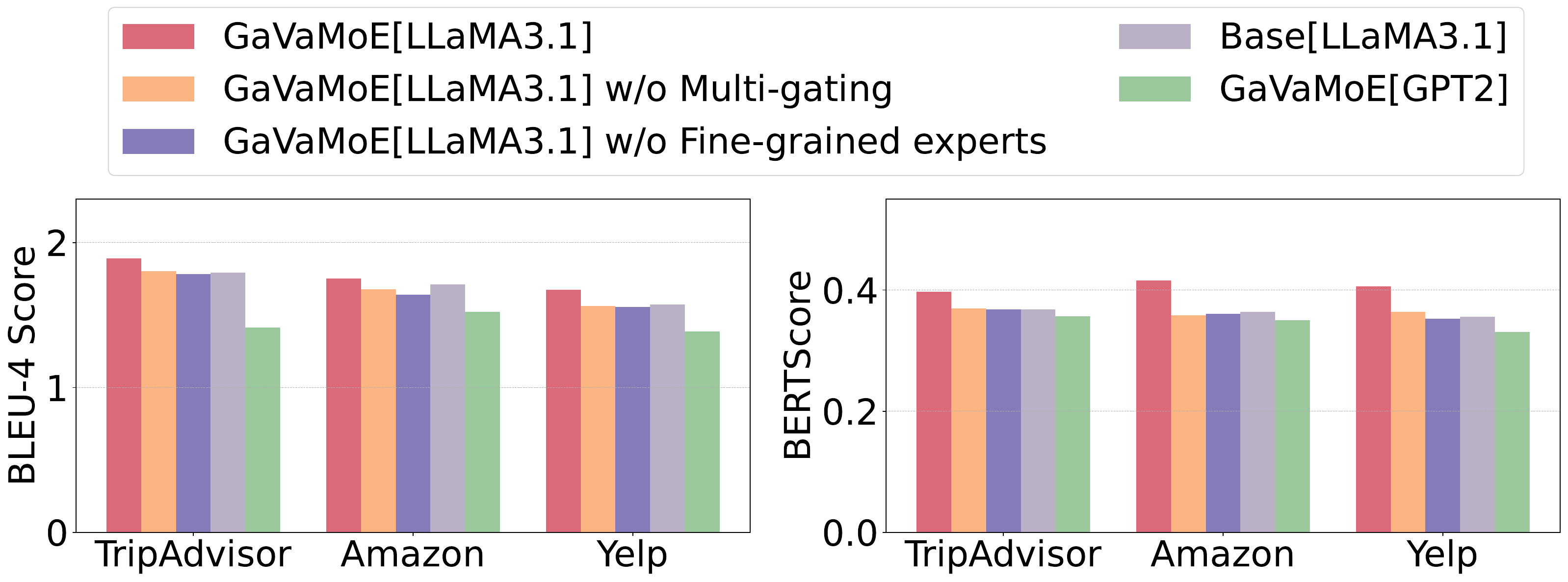} 
    \caption{Ablation study results comparing GaVaMoE variants across TripAdvisor, Amazon, and Yelp datasets.}
    \label{fig:ablation_study}
\end{figure}
\subsubsection{Ablation Study}
To further evaluate the efficacy of GaVaMoE's components, we conducted a comprehensive ablation study across three diverse datasets. We compared the following variants: 
i) \textit{GaVaMoE[LLaMA3.1]}: Our full model that uses LLaMA3.1 as the backbone, incorporating both the multi-gating mechanism and the enhanced MoE architecture into LLaMA3.1's transformer layers.
ii) \textit{GaVaMoE[LLaMA3.1] w/o Multi-gating}: A variant that retains the MoE architecture but removes the multi-gating mechanism.
iii) \textit{GaVaMoE[LLaMA3.1] w/o Fine-grained experts}: A variant that retains the multi-gating mechanism but removes the fine-grained experts.
iv) \textit{Base[LLaMA3.1]}: A baseline version using only the LLaMA3.1 model where the FFN layer acts as a single expert, representing the scenario of expert ensemble without the multi-gating mechanism and MoE architecture.
v) \textit{GaVaMoE[GPT2]}: A variant that employs GPT2 as the backbone architecture, where we integrate our multi-gating mechanism and MoE components into GPT2's transformer layers for explanation generation.
These ablations examine each component's impact and GaVaMoE's backbone choices.

Figure \ref{fig:ablation_study} shows ablation study results across three datasets, comparing BLEU-4 and BERTScore metrics. Key findings include:
(1) \textbf{Effectiveness of Multi-gating Mechanism and Fine-grained Experts}: GaVaMoE[LLaMA3.1] outperformed variants without these components. On Amazon, it scored 1.75 BLEU-4, vs. 1.66 and 1.64 for variants lacking multi-gating and fine-grained experts, respectively, showing their importance in improving explanation quality.
(2) \textbf{Complementary Effects of Architectural Components}: Full GaVaMoE[LLaMA3.1] significantly improved over Base[LLaMA3.1], with BERTScore increases of 5.5\%, 2.3\%, and 6.5\% on TripAdvisor, Amazon, and Yelp. This highlights the combined advantage of multi-gating for precise routing and the enhanced MoE structure for specialized processing.
(3) \textbf{Base Model Independence}: Interestingly, GaVaMoE[GPT2] showed competitive performance, outperforming most baseline models except XRec \cite{ma2024xrec}. This indicates that our multi-gating and MoE components provide substantial improvements even with smaller, less advanced language models.
(4) \textbf{Scalability}: GaVaMoE[GPT2] maintained strong performance across different base model sizes, with only slightly lower BLEU-4 scores than GaVaMoE[LLaMA3.1] (e.g., 1.52 vs. 1.75 on Amazon), demonstrating the architecture's adaptability.
\begin{table*}[]
\resizebox{\textwidth}{!}{
\begin{tabular}
{@{}cccccccccc@{}}
\toprule
Router & BLEU-1  & BLEU-4 & ROUGE-1 & ROUGE-L & DISTINCT-1 & DISTINCT-2 & BERTScore-P & BERTScore-R & BERTScore-F \\ \midrule
\multicolumn{10}{c}{TripAdvisor}                                                                            \\ \midrule
2      & 17.270 & 1.735 & 25.314 & 18.392 & 20.466    & 65.332    & 0.301       & 0.296       & 0.263       \\
3      & \textbf{21.142} & \textbf{1.891} & \textbf{25.465} & \textbf{18.118} & \textbf{22.784}    & \textbf{68.398}    & \textbf{0.397}       & \textbf{0.392}       & \textbf{0.395}       \\
4      & 19.980 & 1.862 & 25.040 & 18.192 & 21.284    & 67.190    & 0.313       & 0.299       & 0.285       \\
5      & 16.038 & 1.748 & 24.886 & 18.034 & 20.130    & 66.036    & 0.290       & 0.259       & 0.240       \\ \midrule
\multicolumn{10}{c}{Amazon (Movie \& TV)}                                                                    \\ \midrule
2      & \textbf{21.370} & \textbf{1.751} & \textbf{25.445} & \textbf{17.747} & \textbf{21.254}    & \textbf{65.784}    & \textbf{0.416}       & \textbf{0.410}       & \textbf{0.412}       \\
3      & 20.547 & 1.653 & 24.971 & 17.537 & 20.370    & 64.444    & 0.409       & 0.398       & 0.401       \\
4      & 18.224 & 1.498 & 23.697 & 16.975 & 19.470    & 63.454    & 0.388       & 0.370       & 0.373       \\
5      & 16.047 & 1.326 & 22.067 & 16.644 & 17.201    & 61.241    & 0.360       & 0.334       & 0.347       \\ \midrule
\multicolumn{10}{c}{Yelp}                                                                                   \\ \midrule
2      & 17.465 & 1.521 & 21.072 & 15.271 & 17.113    & 60.074    & 0.311       & 0.290       & 0.304       \\
3      & \textbf{19.586} & \textbf{1.616} & \textbf{23.259} & \textbf{17.416} & \textbf{19.483}    & \textbf{63.847}   & \textbf{0.406}       & \textbf{0.374}       & \textbf{0.381}       \\
4      & 16.225 & 1.396 & 22.695 & 16.876 & 17.471    & 61.455    & 0.369       & 0.351       & 0.352       \\
5      & 15.013 & 1.284 & 21.088 & 15.724 & 16.211    & 60.145    & 0.357       & 0.326       & 0.339       \\ \bottomrule
\end{tabular}
}
\caption{Performance comparison of GaVaMoE with different numbers of gates (2-5) across three datasets. Results show that \textbf{3 gates} achieves optimal performance for TripAdvisor (BLEU-4: 1.891) and Yelp (BERTScore-F: 0.381), while \textbf{2 gates} performs best for Amazon (BLEU-4: 1.751, BERTScore-F: 0.412), demonstrating that optimal gate numbers vary across different datasets and tasks.}
\label{gatestudy}
\end{table*}
\begin{figure}[]
    \centering
    \includegraphics[width=\columnwidth]{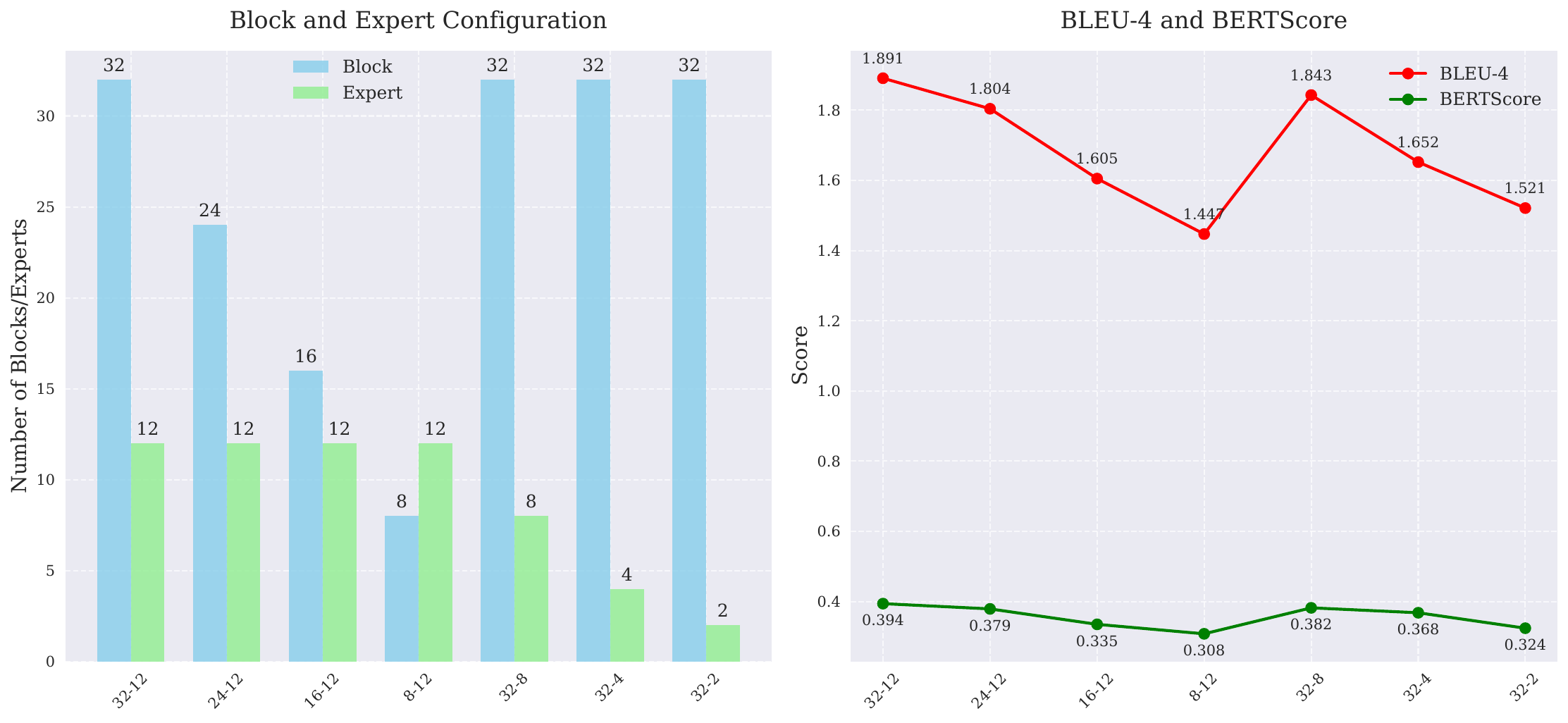} 
    \caption{Impact of block and expert configurations on GaVaMoE performance. Left: Bar chart showing different combinations of transformer blocks (blue) and experts (green) tested in the experiments. Right: Line plot demonstrating the corresponding BLEU-4 and BERTScore metrics for each configuration, where configurations are denoted as "blocks-experts" (e.g., "32-12" represents 32 blocks with 12 experts).}
    \captionsetup{skip=0pt}
    \label{fig:block_exp}
\end{figure}

\subsubsection{Analysis of Multi-gating Mechanism}

To evaluate the impact of our multi-gating mechanism, we varied the number of gates from 2 to 5 across three datasets, while maintaining the backbone architecture of LLaMA3.1-8B which consists of 32 transformer blocks. Within each block, we implement 12 experts with 2 experts being activated for processing at each pass. Table \ref{gatestudy} shows distinct optimal gate configurations: 3 gates for TripAdvisor (BLEU-4: 1.891, BERTScore-F: 0.395) and Yelp (BLEU-4: 1.616, BERTScore-F: 0.381), while Amazon achieves best performance with 2 gates (BLEU-4: 1.751, BERTScore-F: 0.412).

Our key findings include: (1) \textbf{Dataset-Dependent Performance}: Different datasets benefit from different routing granularities. Amazon, with more structured user preferences in the Movie \& TV domain, performs best with 2 gates. In contrast, TripAdvisor and Yelp, covering diverse business categories, require 3 gates to capture complex preference patterns. This suggests that the optimal number of gates should be tailored to the inherent complexity and diversity of the domain.
(2) \textbf{Performance Trade-offs}: Our experiments demonstrate a clear trend where performance peaks at an optimal gate number (2 or 3) and then declines with additional gates, suggesting that excessive routing paths can fragment the learning process and lead to suboptimal training of both gates and their associated experts. This observation emphasizes the importance of finding a balance between routing flexibility and training stability.
\subsubsection{Analysis of Experts and Blocks}
\begin{figure}[]
    \centering
    \begin{subfigure}[t]{0.5\columnwidth}
        \centering
        \includegraphics[width=\columnwidth]{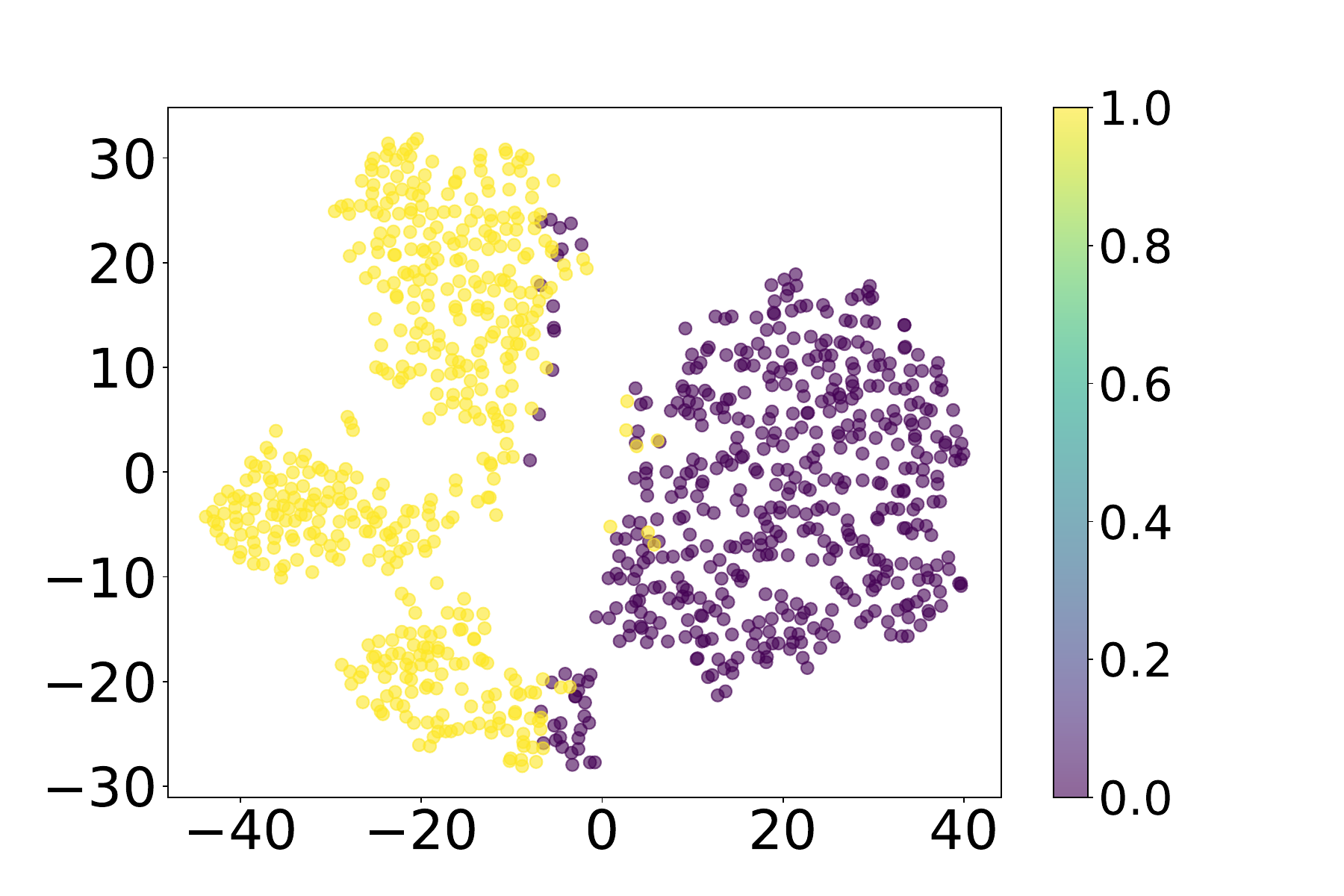}
        \label{fig:cluster2}
    \end{subfigure}
    \hspace{-10pt}  
    \begin{subfigure}[t]{0.5\columnwidth}
        \centering
        \includegraphics[width=\columnwidth]{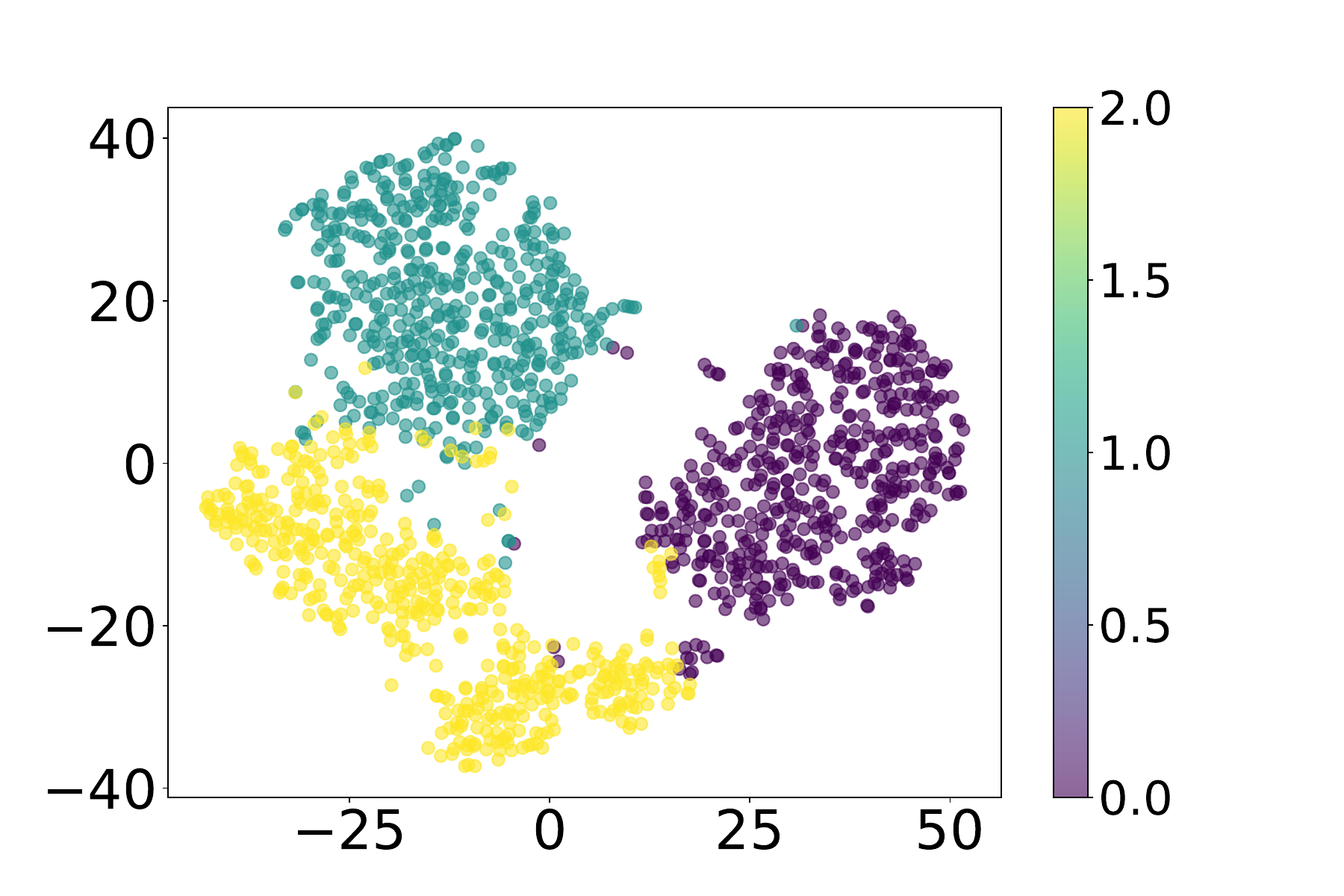}
        \label{fig:cluster3}
    \end{subfigure}

    \vspace{-1em}  

    \begin{subfigure}[t]{0.5\columnwidth}
        \centering
        \includegraphics[width=\textwidth]{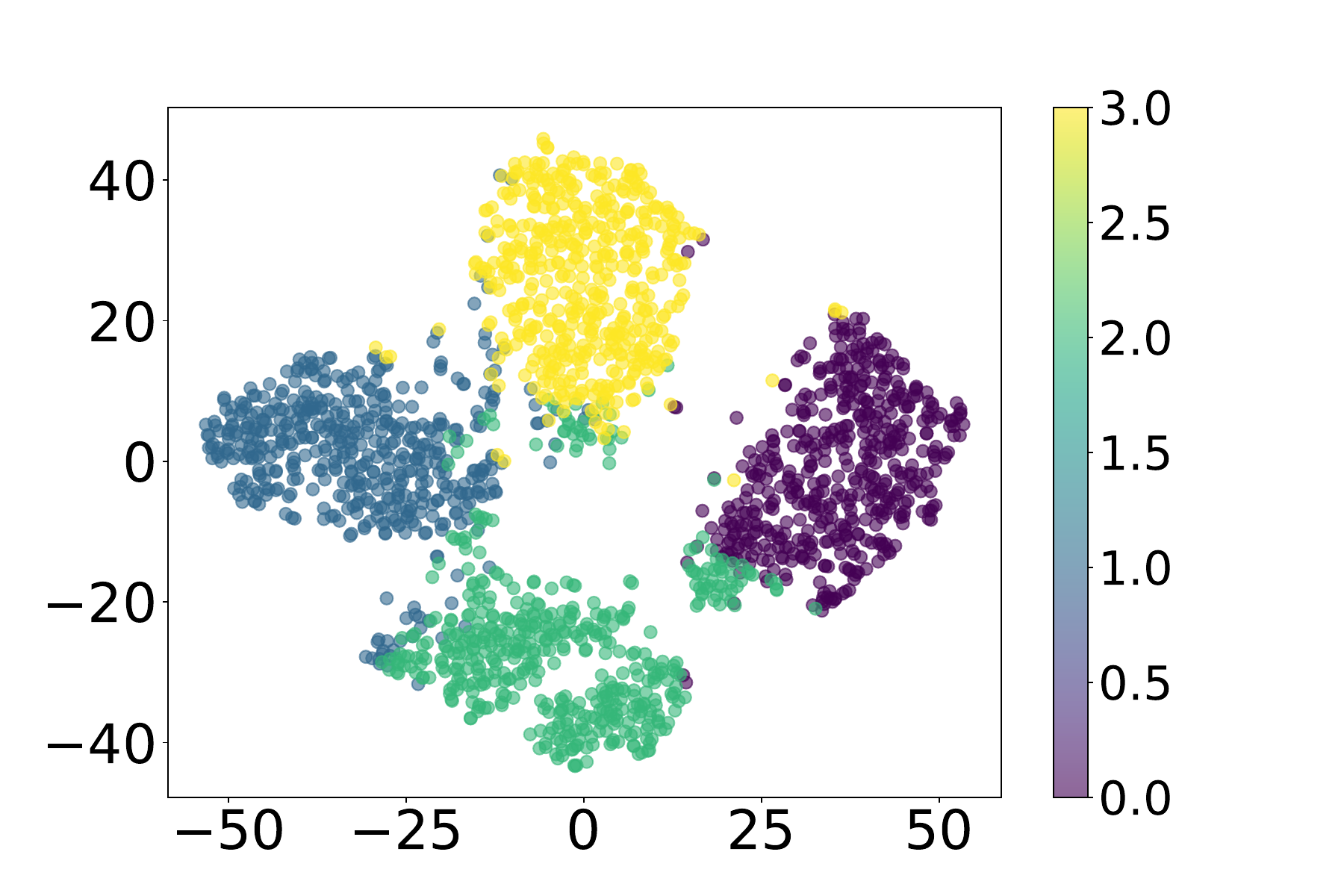}
        \label{fig:cluster4}
    \end{subfigure}
    \hspace{-10pt}  
    \begin{subfigure}[t]{0.5\columnwidth}
        \centering
        \includegraphics[width=\columnwidth]{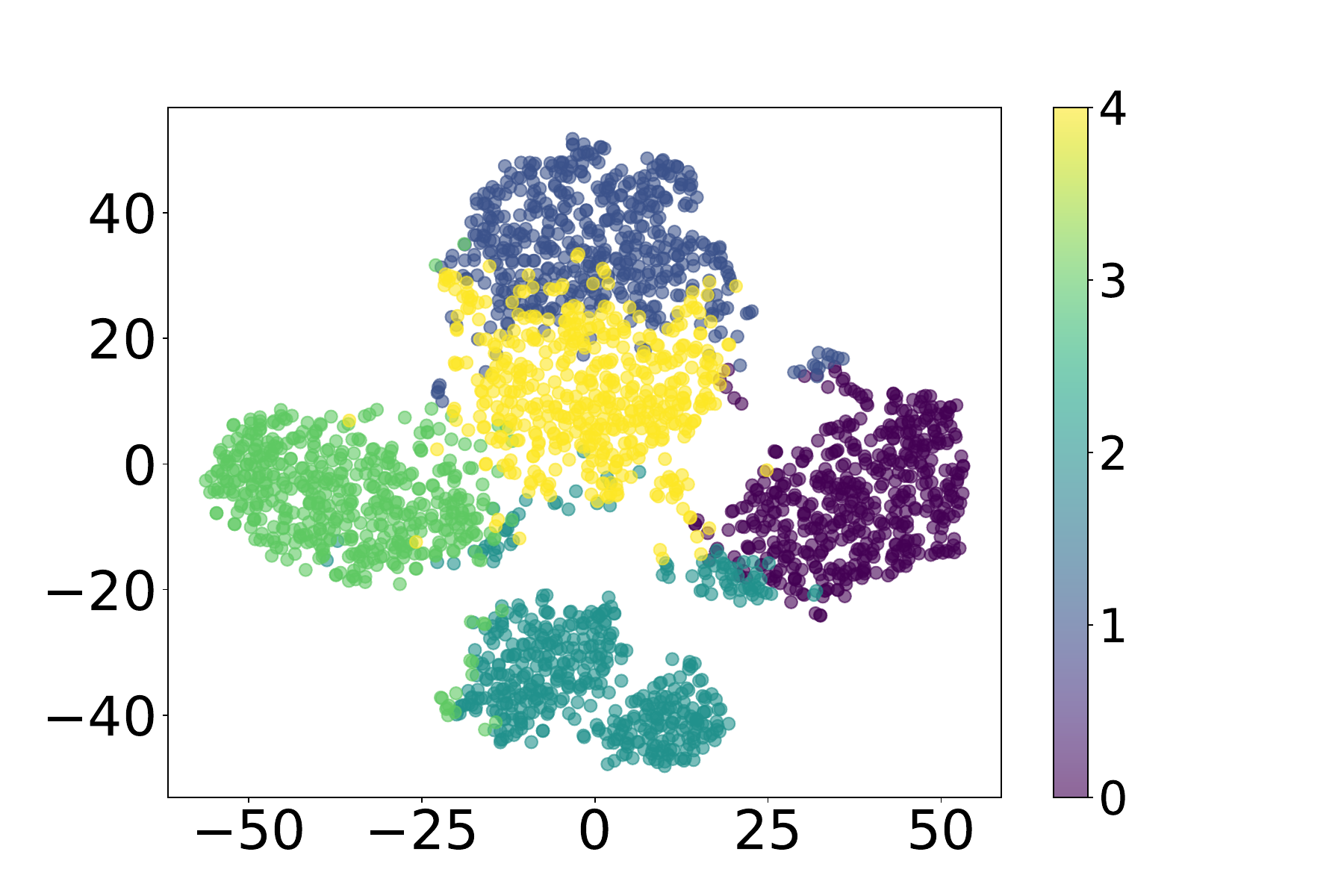}
        \label{fig:cluster5}
    \end{subfigure}
    \vspace{-1em}
    \caption{Visualization of latent space distributions in GaVaMoE for different numbers of user clusters.}
    \label{fig:clusters2-5}
\end{figure}
To investigate GaVaMoE's architectural sensitivity, we conducted experiments varying the numbers of blocks and experts (Figure \ref{fig:block_exp}). Our results show that reducing the number of blocks in LLaMA-3.1-8B from 32 to 8 leads to a significant drop in BLEU-4 scores (1.891 to 1.447), indicating the importance of deep layers in capturing rich semantic information for high-quality explanations. Similarly, decreasing experts from 12 to 2 while maintaining 32 blocks results in BLEU-4 score degradation from 1.891 to 1.521, demonstrating that sufficient expert specialization is crucial for modeling diverse user preferences and generating personalized explanations effectively. The observed performance improvements can be attributed to two key factors. More GaVaMoE blocks enable hierarchical feature extraction, capturing both low-level patterns and high-level semantic relationships in user-item interactions. Meanwhile, a larger number of experts facilitates finer-grained specialization, with each expert focusing on specific user preference patterns, thereby enabling more precise and personalized explanations.
\subsubsection{Analysis of Data Sparsity}
\begin{figure}[]
    \centering
    \includegraphics[width=\columnwidth]{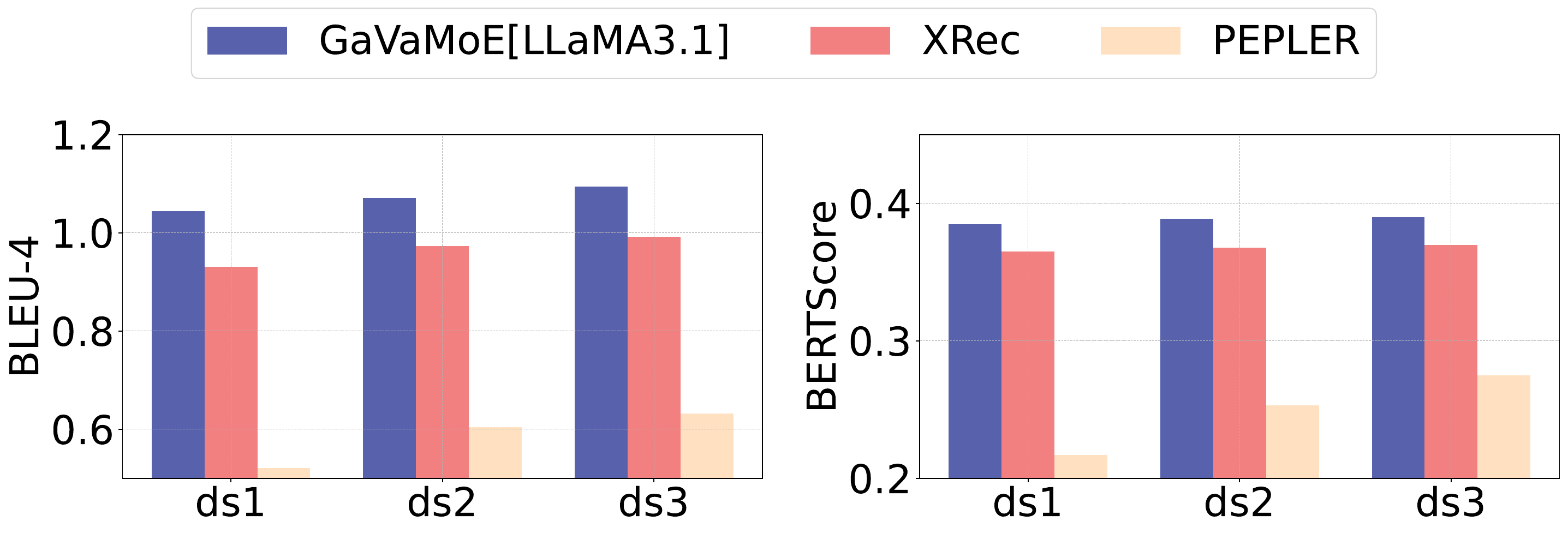} 
    \caption{Performance comparison across different sparsity levels (ds1, ds2, ds3). GaVaMoE demonstrates consistent superiority and stability across all sparsity levels.}
    \label{fig:example_image2}
\end{figure}

To evaluate GaVaMoE's robustness against data sparsity, we conducted a comprehensive evaluation of GaVaMoE's performance across varying levels of user-item interaction sparsity. Following methodologies from \cite{ma2024xrec} and \cite{cai2022pevae}, we divided the test data into three equal subsets (ds1, ds2, ds3) based on user appearance frequency in the training data. This division creates a spectrum of sparsity, with ds1 containing the most frequent users and ds3 the least frequent, allowing us to assess model performance under increasingly sparse conditions. As shown in Figure \ref{fig:example_image2}, GaVaMoE consistently outperforms XRec and PEPLER across all sparsity levels in terms of BLEU-4 and BERTScore metrics. Notably, GaVaMoE's performance remains stable from ds1 to ds3, indicating its effectiveness in handling sparse data scenarios. Furthermore, the case studies presented in Appendix \ref{app:casestudy} provide qualitative evidence of GaVaMoE's superior performance in both limited and sufficient user-item interaction scenarios.

The strong performance of GaVaMoE under sparse conditions can be attributed to two key mechanisms. First, the VAE-GMM component effectively clusters users based on behavioral patterns, creating meaningful user groups even with limited data. As visualized in Figure \ref{fig:clusters2-5}, this clustering enables knowledge transfer from users with rich interaction histories to similar users with sparse data within the same cluster. Second, our multi-gating mechanism dynamically routes user-item pairs to specialized experts, ensuring personalized explanations by leveraging collective cluster knowledge rather than relying solely on individual user histories.
\section{Conclusion}
In this paper, we introduced GaVaMoE, a novel Gaussian-Variational Gated Mixture of Experts framework for explainable recommendation systems. GaVaMoE addresses critical challenges in existing LLM-based ER systems through two key innovations: (1) a rating reconstruction module employing VAE with GMM for capturing complex user-item collaborative preferences, and (2) a multi-gating mechanism coupled with expert models for generating personalized explanations. Our approach enhances collaborative preference modeling, improves personalization, and effectively handles data sparsity. Extensive experiments on three real-world datasets demonstrate that GaVaMoE significantly outperforms state-of-the-art methods across multiple metrics, including explanation quality, personalization, and consistency. Notably, GaVaMoE exhibits robust performance in sparse data scenarios, maintaining high-quality explanations even for users with limited historical data. 
\section*{Limitations}
GaVaMoE introduces a multi-gating mechanism with multiple experts to model user preferences, which works well in controlled environments. However, in real-world applications, it may not be practical to deploy an excessive number of experts to capture the full diversity of user behaviors. As the number of experts increases, the complexity of the model also grows, leading to higher computational costs and potential difficulties in scaling. Additionally, while the multi-gating mechanism is effective in routing user-item pairs to specialized experts, there may be diminishing returns in terms of performance when the number of experts becomes too large relative to the available data. In real-world scenarios, user preferences are often complex but not infinitely diverse, and it may be more efficient to adopt a more compact expert structure without sacrificing performance.
\bibliography{anthology}
\clearpage
\appendix
\section{Related Works}
\textbf{Explainable Recommendation System} Explainable recommendation provides explanations that clarify why certain items are recommended, thereby enhancing the transparency and persuasiveness of the recommendation systems \cite{zhang2020explainable,tintarev2015explaining}.
Explainable recommendations can be presented in various format, such as predefined templates \cite{tan2021counterfactual, zhang2014explicit, li2020generate}, reasoning rules \cite{chen2021neural, shi2020neural}. However, these methods are expensive to maintain and fail to produce diverse, personalized explanations, resulting in poor generalization. Related works of Explainable recommendation based on item features \cite{he2015trirank, wang2018tem}, knowledge graph paths \cite{ai2018learning, fu2020fairness, xian2019reinforcement}, and ranked text \cite{li2021extraexplanationrankingdatasets, li2021extra} also similarly face the problem of low generalization and poor personalization. 
To address these challenges, more and more explainable recommendation systems based on natural language processing techniques have been studied. These works focus on using generative models to directly obtain personalized explanations. NRT \cite{li2017neural} simultaneously performs accurate score prediction while generating high-quality summarized prompts by integrating user and item latent factors. Co-Attentive Multi-Task Learning (CAML ) \cite{chen2019co} integrates a multi-task learning mechanism and adopts the joint attention proposed in \cite{tay2018multi}. PETER\cite{li2021personalized} employs a small, unpretrained Transformer, connecting user and item IDs with generated text through a designed context prediction task for personalized text generation. 

Recently, research on LLM-based ER systems has gained significant attention. ReXPlug is an end-to-end explainable recommendation framework that generates high-quality personalized natural language reviews for users by plugging in and utilizing a plug-and-play language model. PEPLER \cite{li2023personalized} utilizes a pretrained language model GPT2 \cite{radford2019language} to generate explainable recommendations by incorporating user and item ID vectors into prompts. LLM2ER \cite{yang2024fine} utilizes a personalized prompt learning module to match user preference and  fine-tunes the model with reinforcement learning using two innovative explainability quality reward models to generate explanations. Xrec \cite{ma2024xrec} seamlessly integrates the capabilities of large language models with a graph-based collaborative filtering paradigm to generate text. To address the low-quality issues caused by hallucination in large language models, PEVAE \cite{cai2022pevae} employs personalized variational autoencoders to tackle specific quality concerns, while PRAG \cite{xie2023factual} enhances performance by incorporating retrieval augmentation to resolve certain low-quality issues. \\
\textbf{Mixture of Experts} Mixture of Experts (MoE) technology was first introduced by \cite{jacobs1991adaptive}, where different experts are used to complete various subtasks. 
With the development of deep learning, the Transformer \cite{vaswani2017attention} has been widely used in various natural language processing tasks. In the process, numerous studies have attempted to replace the feed-forward neural network (FFN) layer within Transformer with MoE.
Sparsely-Gated MoE \cite{shazeer2017outrageously} introduces a sparsely gated mixture of experts layer, which contains multiple feed-forward layers and scales up to 137B parameters. This MoE implementation is applied convolutely between stacked LSTM \cite{hochreiter1997long} layers. GShard \cite{lepikhin2020gshard} is a pioneer in scaling MoE language models to ultra-large sizes using learnable top-2 or top-1 routing strategies. Switch Transformer \cite{fedus2022switch} introduces the concept of expert capacity and simplifies the expert routing algorithm, summarizing a set of MoE training experiences. The ST-MoE \cite{zoph2022st} model is a stable and transferable sparse expert model that addresses training instability issues by introducing router z-loss. With the development of large language models, researchers have made significant innovations in enhanced MoE. Mixtral 8x7B \cite{jiang2024mixtral} is a sparse mixture of experts language model with 8 feed-forward blocks, achieving performance comparable to a 70B parameter model across multiple benchmarks. 
DeepSeekMoE \cite{dai2024deepseekmoe} innovatively proposes fine-grained expert segmentation and shared expert isolation strategies to address the problems of knowledge mixing and redundancy, enhancing model performance.
\section{Details on Training and Inference}
\label{sec:details_train}
\subsection{Implementation Detail}
\label{sec:imple_detail}
GaVaMoE employs a two-layer Transformer encoder and a two-layer MLP for the decoder in the VAE. The latent space size is 128, with user and item embeddings of 768. We set batch size to 4096, learning rate to 1e-5, $\beta$ to 0.1 and 30 training epochs. For Stage 2 (Explanation Generation), we set $\alpha$ to 0.1, batch size to 1, learning rate to 3e-5. The training process utilizes AdamW with a gradient accumulation of 8 and a clipping norm of 0.3. GaVaMoE uses LLaMA3.1-8B \cite{dubey2024llama} with 32 GaVaMoE blocks. We replaced the FFN in LLaMA3.1-8B with MoE, setting the hidden size $d$ to 4096 and the number of experts to 6. Then, by setting $r$ to 2, each MoE was split, reducing the hidden size by half while doubling the number of experts. Additionally, we activated 2 experts per pass, with the activation parameter denoted as $k$. As a result, the final fine-grained configuration has a hidden size of 2048 and 12 experts. The number of gates matches user clusters, and explanation generation training lasts 3 for epochs.
\subsection{Compared Methods}
We compare our model’s performance against the following competiable baselines in explainable recommendation:
\begin{itemize}
    \item NETE\cite{li2020generate}: presents a gated fusion recurrent unit that leverages neural templates to generate high-quality, explainable text for recommendation systems.
    \item PETER \cite{li2021personalized}: utilizes Transformer to integrate user and item IDs into generated explanations.
    \item PEPLER \cite{li2023personalized}: adopts a pre-trained GPT-2 model with prompt tuning for explanation generation.
    \item PEVAE \cite{cai2022pevae}: extends hierarchical VAEs to address data sparsity in LLM-based recommendation systems, aligning with our focus on sparse interaction scenarios.
    \item XRec \cite{ma2024xrec}: generates explanatory texts by integrating high-order collaborative signals encoded by graph neural networks and the text generation capabilities of LLMs.
\end{itemize}

These baselines were carefully selected to represent the progression of explainable recommendation systems, from template-based approaches (NETE) to advanced LLM-based models (XRec).

\subsection{Evaluation Metrics}
To comprehensively evaluate GaVaMoE's performance, we employ metrics assessing three key dimensions: explanation quality, personalization, and consistency. For explanation quality, we use BLEU \cite{papineni2002bleu} (BLEU-1 and BLEU-4) and ROUGE \cite{lin2004rouge} (ROUGE-1 and ROUGE-L) to assess fluency, grammatical correctness, and content overlap. For personalization, we employ Distinct-1 and Distinct-2, evaluating unigram and bigram diversity in generated text. For consistency, we utilize BERTScore \cite{zhang2019bertscore}, capturing deeper semantic similarities using contextual embeddings. While BLEU and ROUGE are widely used, they have limitations in capturing true semantic information, relying primarily on n-gram overlap. BERTScore addresses this limitation with a more sophisticated approach to evaluating semantic consistency. This diverse set of metrics allows for a comprehensive evaluation of GaVaMoE's performance across various aspects of explanation generation, assessing not only linguistic quality but also personalization and semantic consistency.
\section{ELBO Formulation in GaVaMoE}
\label{app:elbo}
In GaVaMoE, our objective is to maximize the likelihood of user-item pairs, equivalent to maximizing $\log p(x)$. Adopting Jensen's inequality, we derive the variational evidence lower bound (ELBO) as follows:
\begin{equation}
\begin{aligned}
\mathcal{L}_{ELBO}(x) &= \mathbb{E}_{q(z,c|x)} \left[\log \frac{p(c,z,x)}{q(z,c|x)}\right] \\
&= \mathbb{E}_{q(z,c|x)} \big[\log p(x|z) + \\
&\log p(z|c) + \log p(c) \big] \\
&\quad - \mathbb{E}_{q(z,c|x)} \big[\log q(z|x) + \log q(c|x)\big] \\
&= \mathbb{E}_{q(z,c|x)} \left[\log p(x|z) + \log \frac{p(z,c)}{q(z,c|x)}\right]
\end{aligned}
\end{equation}

The first term represents the reconstruction loss, allowing the model to discover deep level user-item interactions. The second term calculates the Kullback-Leibler divergence between the mixed Gaussian prior distribution \(q(z|x)\) and the variational posterior \(p(z,c)\). To learn more disentangled representations, according to \(\beta\)-VAE\cite{higgins2017beta} approach, we added a hyperparameter \(\beta\) to the KL divergence term, rewriting the formula as follows:
\begin{equation}
\begin{aligned}
\mathcal{L}_{\text{ELBO}}(x, \beta) 
&= \mathbb{E}_{q(z, c | x)} \left[ \log p(x | z) \right] \\
&\quad - \beta \cdot \text{KL}\left(q(z, c | x) \| p(z, c)\right)
\end{aligned}
\label{elbo_equ}
\end{equation}

\noindent where $\beta$ balances the reconstruction loss and the KL regularization term. The KL regularization term is derived as:

\begin{equation}
\begin{aligned}
& \text{KL}(q(z, c | x) \| p(z, c)) \\
&= -\frac{1}{2} \sum_{c=1}^{K} \gamma_{i,c} \sum_{d=1}^{D} 
\left[
    \log (\bar{\sigma}_{c,d})^2 
    + \left( \frac{\sigma_{i,d}}{\bar{\sigma}_{c,d}} \right)^2
\right] \\
&\quad + \sum_{c=1}^{K} \gamma_{i,c} 
\log \frac{\pi_c}{\gamma_{i,c}}
+ \frac{1}{2} \sum_{d=1}^{D} 
\left[
    \log (\sigma_{i,d})^2
\right] \\
&\quad - \frac{1}{2} \sum_{c=1}^{K} \gamma_{i,c} \sum_{d=1}^{D} 
\left[
    \frac{(\mu_{i,d} - \bar{\mu}_{c,d})^2}{\bar{\sigma}_{c,d}^2}
\right] 
+ \frac{1}{2}
\end{aligned}
\label{eq:elbo_loss}
\end{equation}

In this formulation, $K$ represents the number of clusters, $D$ is the dimensionality of the latent space, $\gamma_{i,c}$ is the probability that the $i$-th sample belongs to cluster $c$, and $\pi_c$ is the prior probability of cluster $c$. The parameters $\mu_{i,d}$ and $\sigma_{i,d}$ are the mean and standard deviation of the $i$-th sample in the d-th dimension, while $\bar{\mu}{c,d}$ and $\bar{\sigma}{c,d}$ are the mean and standard deviation of cluster $c$ in the d-th dimension.

By optimizing this ELBO, GaVaMoE can effectively learn a structured latent space that captures both user-item interactions and user clustering information, which is crucial for the subsequent multi-gating mechanism and personalized explanation generation.

\begin{table}[!ht]
\centering
\scalebox{0.9}{
\begin{tabular}{cp{7cm}}
\toprule
\textbf{Score} & \textbf{Criteria Description} \\
\midrule
\multicolumn{2}{l}{\textbf{Personalization (P)}} \\
5 & Explanation perfectly aligns with user preferences, demonstrating deep understanding of individual tastes. \\
4 & Explanation shows strong alignment with user preferences, with minor misalignments. \\
3 & Explanation moderately aligns with user preferences, with some noticeable misalignments. \\
2 & Explanation shows weak alignment with user preferences, with significant misalignments. \\
1 & Explanation shows no alignment with user preferences. \\
\midrule
\multicolumn{2}{l}{\textbf{Text Quality (Q)}} \\
5 & Text is exceptionally clear, coherent, and well-structured, with no errors. \\
4 & Text is clear and coherent, with minimal structural issues or errors. \\
3 & Text is generally clear but has some issues with coherence or structure. \\
2 & Text has notable issues with clarity, coherence, or structure. \\
1 & Text is unclear, incoherent, or poorly structured. \\
\midrule
\multicolumn{2}{l}{\textbf{User Satisfaction (U)}} \\
5 & Explanation is highly engaging, likely to prompt immediate user interaction or action. \\
4 & Explanation is engaging and likely to prompt user interaction. \\
3 & Explanation is somewhat engaging, may prompt user interaction. \\
2 & Explanation is unlikely to prompt user interaction. \\
1 & Explanation is likely to discourage user interaction. \\
\bottomrule
\end{tabular}
}
\caption{Scoring Criteria for Human Evaluation}
\label{tab:scoring_criteria}
\end{table}

\begin{table*}[!htb]
\centering
\label{casestudy}
\begin{tabular}{@{}p{0.47\textwidth}|p{0.47\textwidth}@{}}
\toprule
\multicolumn{1}{c|}{\textbf{Case 1: Limited user-item interactions}} & \multicolumn{1}{c}{\textbf{Case 2: Sufficient user-item interactions}} \\
\midrule
\rowcolor[gray]{0.9}
\multicolumn{2}{l}{\textbf{Ground Truth}} \\
\midrule
\textbf{Text:} The restaurant offers authentic Thai cuisine in a cozy setting, with excellent pad thai. & \textbf{Text:} A tech enthusiast's dream smartwatch with advanced fitness tracking, seamless integration with other devices, and impressive battery life. \\
\textbf{Features:} Thai, authentic, pad thai, cozy & \textbf{Features:} tech, smartwatch, fitness tracking, battery life, device integration \\
\midrule
\rowcolor[gray]{0.9}
\multicolumn{2}{l}{\textbf{Generated Explanations}} \\
\midrule
\textbf{PETER:} This is a good restaurant with tasty food. & \textbf{PETER:} This is a good smartwatch with many features. \\
\textbf{PEPLER:} The food here is delicious and worth trying. & \textbf{PEPLER:} A high-tech smartwatch that tracks fitness and has good battery. \\
\textbf{XRec:} A nice Thai place with decent food, but the ambiance needs improvement. & \textbf{XRec:} An advanced smartwatch with fitness features and decent battery life. The integration could be better. \\
\textbf{GaVaMoE:} This authentic Thai restaurant serves exceptional pad thai in a welcoming atmosphere, staying true to traditional flavors. & \textbf{GaVaMoE:} Perfect for tech enthusiasts, this smartwatch excels with its comprehensive fitness tracking capabilities, seamless ecosystem integration, and long-lasting battery life. The features align well with your interest in connected devices and health monitoring. \\
\bottomrule
\end{tabular}
\caption{Comparison of explanations generated by baselines and GaVaMoE under different user interaction scenarios. Case 1 represents a user with limited interaction history, while Case 2 represents a user with sufficient interactions.}
\label{tab:casestudy}
\end{table*}

\section{Human scoring Criteria}
\label{sec:scoring}

This guide provides detailed instructions for evaluating generated explanations on a 1-to-5 scale across three dimensions: Personalization, Text Quality, and User Satisfaction:

\begin{itemize}
    \item \textbf{Personalization (P):} Consider the user's past interactions, stated preferences, and behavioral patterns when assessing alignment. Look for specific mentions of user-relevant features or experiences.
    \item  \textbf{Text Quality (Q):} Evaluate the explanation's grammatical correctness, logical flow, and appropriate use of language. Consider whether the text would be easily understood by the average user.
    \item \textbf{User Satisfaction (U):} Assess whether the explanation would likely motivate the user to engage further with the recommended item or the platform. Consider factors such as informativeness, persuasiveness, and appeal.
\end{itemize}

A detailed scoring guide with specific criteria for each score level is provided in Table \ref{tab:scoring_criteria}, ensuring consistency and reliability in the human evaluation process.

To ensure comprehensive and precise evaluations, human experts are encouraged to utilize the full range of the scale, including half-point scores (e.g., 3.5) when explanations fall between two score descriptions. This granular approach allows for a more nuanced assessment of the explanations' quality and effectiveness.
\begin{table}[]
\small
\centering
\begin{tabular}{ccccc}
\toprule
Method & P & Q & U & Human Score \\
\midrule
GaVaMoE & \textbf{4.524} & \textbf{4.553} & \textbf{4.541} & \textbf{4.538} \\
XRec & 4.390 & 4.287 & 4.252 & 4.317 \\
PEPLER & 3.766 & 3.652 & 3.773 & 3.734 \\
\bottomrule
\end{tabular}
\caption{The human evaluation results on the Yelp dataset.}
\label{tab:performance_yelp}
\end{table}

\section{Human Evaluation}
To validate GaVaMoE's effectiveness in real-world scenarios, we conducted a human evaluation study using 2,000 randomly sampled entries from the Yelp dataset. We compared GaVaMoE against XRec and PEPLER, employing three professionals to assess the generated explanations based on personalization (P), text quality (Q), and user satisfaction (U). Each criterion was scored on a 1-to-5 scale, with details provided in Table \ref{tab:scoring_criteria}. We computed a composite Human-Score using a weighted formula:
\begin{equation}
    \text{Human-Score} =  \lambda_{p}P_{score} + \lambda_{q}Q_{score} + \lambda_{u}U_{score}
\end{equation}
Where \(\lambda_p = 0.4\), \(\lambda_q\), and \(\lambda_u = 0.3\). The scores P, Q, and U represent the metrics for personalization, text quality, and user satisfaction, respectively.

Table \ref{tab:performance_yelp} presents these results, showing that GaVaMoE consistently outperformed baselines across all metrics. GaVaMoE achieved the highest overall Human-Score of 4.538, significantly surpassing XRec (4.317) and PEPLER (3.734). These human evaluation results corroborate our computational findings, demonstrating GaVaMoE's ability to generate high-quality, personalized explanations that resonate with users in real-world scenarios.

\section{Case Study}
\label{app:casestudy}
To further evaluate GaVaMoE's effectiveness in generating personalized and relevant explanations, we conducted a detailed case study comparing our model's performance with baseline methods under different user interaction scenarios. Table \ref{tab:casestudy} presents a comparative analysis of explanations generated by GaVaMoE and baseline methods under two distinct scenarios: limited and sufficient user-item interactions. GaVaMoE demonstrates superior performance across both cases, effectively addressing the challenges of data sparsity and personalization. 

In Case 1 (limited interactions), where the user has minimal historical data about restaurant preferences, baseline methods struggle to generate informative explanations. PETER and PEPLER produce overly generic statements ("good restaurant", "delicious food") that fail to capture specific attributes. While XRec identifies the cuisine type, it introduces potentially inaccurate information about the ambiance. In contrast, GaVaMoE successfully captures key features ("authentic Thai", "pad thai", "welcoming atmosphere") and generates a coherent, accurate explanation despite the sparse interaction data. This demonstrates GaVaMoE's robustness in handling data sparsity through its effective clustering and knowledge transfer mechanisms.

Case 2 (sufficient interactions) involves a tech product review where rich user interaction history is available. Here, GaVaMoE leverages the comprehensive user preference data to generate highly personalized explanations. While baseline methods manage to identify basic features (e.g., "smartwatch", "fitness tracking"), they fail to establish meaningful connections with user preferences. GaVaMoE's explanation stands out by not only covering all key product features but also connecting them to the user's demonstrated interests in technology and health monitoring. The explanation reflects a deep understanding of both product attributes and user preferences, achieved through GaVaMoE's sophisticated VAE-GMM preference modeling and multi-gating mechanism.

Across both scenarios, GaVaMoE demonstrates consistent superiority in three key aspects: (1) Feature Coverage - capturing comprehensive item characteristics regardless of data sparsity, (2) Accuracy - avoiding the introduction of unsubstantiated information, and (3) Personalization - generating explanations that meaningfully reflect user preferences and interaction patterns. These observations validate GaVaMoE's effectiveness in addressing the dual challenges of data sparsity and personalization in explainable recommendation systems.

\end{document}